\documentstyle[aps,prl,preprint]{revtex}
\def\fun#1#2{\lower3.6pt\vbox{\baselineskip0pt\lineskip.9pt
        \ialign{$\mathsurround=0pt#1\hfill##\hfil$\crcr#2\crcr\sim\crcr}}}

\begin{document}

\date {IMPERIAL/TP/93-94/58 $~~~~$   DART-HEP-94/06 $~~~~$  September 1994}

\vspace{0.5in}
\title{Thermal Phase Mixing During First Order Phase Transitions}

\vspace{1.cm}

\author{Julian Borrill$^{1)}$ and Marcelo Gleiser\thanks{NSF Presidential
Faculty Fellow}$^{2)}$}

\vspace{1.0cm}

\address{ $^{1)}$ Blackett Laboratory, Imperial College\\
Prince Consort Road, London SW7 2BZ, UK}

\address{ $^{2)}$ Department of Physics and Astronomy, Dartmouth College\\
Hanover, NH 03755, USA}

\maketitle

\vspace{1.cm}

\begin{abstract}
\baselineskip 16pt

The dynamics of first order phase transitions are studied in the
context of (3+1)-dimensional scalar field theories. Particular
attention is paid to the question of quantifying the strength of the
transition, and how `weak' and `strong' transitions have different
dynamics. We propose a model with two available low temperature phases
separated by an energy barrier so that one of them becomes metastable
below the critical temperature $T_c$. The system is initially prepared
in this phase and is coupled to a thermal bath. Investigating the
system at its critical temperature, we find that `strong' transitions
are characterized by the system remaining localized within its initial
phase, while `weak' transitions are characterized by considerable
phase mixing. Always at $T_c$, we argue that the two regimes are
themselves separated by a (second order) phase transition, with an
order parameter given by the fractional population difference between
the two phases and a control parameter given by the strength of the
scalar field's quartic self-coupling constant. We obtain a
Ginzburg-like criterion to distinguish between `weak' and `strong'
transitions, in agreement with previous results in (2+1)-dimensions.

\end{abstract}

\def\p{\phi}
\def\P{\Phi}
\def\a{\alpha}
\def\l{\lambda}
\def\s{\sigma}

\newpage

\def\beq{\begin{equation}}
\def\eeq{\end{equation}}
\def\beqa{\begin{eqnarray}}
\def\eeqa{\end{eqnarray}}
\baselineskip 24pt
\section{Introduction}

The fact that the gauge symmetries describing particle interactions
can be restored at high enough temperatures has led, during the past
15 years or so, to an active research program on the possible
implications that this symmetry restoration might have had to the
physics of the very early Universe. One of the most interesting and
popular possibilities is that during its expansion the Universe
underwent a series of phase transitions, as some higher symmetry group
was successively broken into products of smaller groups, up to the
present standard model described by the product $SU(3)_C\otimes
SU(2)_L \otimes U(1)_Y$. Most models of inflation and the formation of
topological (and nontopological defects) are well-known consequences
of taking the existence of cosmological phase transitions seriously
\cite{KT}.

One of the motivations of the present work comes from the possibility
that the baryon asymmetry of the Universe could have been dynamically
generated during a first order electroweak phase transition \cite{EW}.
As is by now clear, a realistic calculation of the net baryon number
produced during the transition is a formidable challenge. We probably
must invoke physics beyond the standard model (an exciting prospect
for most people) \cite{FS}, push perturbation theory to its limits
(and beyond, due to the nonperturbative nature of magnetic plasma
masses that regulate the perturbative expansion in the symmetric
phase), and we must deal with nonequilibrium aspects of the phase
transition. Here we will focus on the latter problem, as it seems to
us to be the least discussed of the pillars on which most baryon
number calculations are built upon. To be more specific, we can
separate the nonequilibrium aspects of the phase transition into two
main subdivisions. If the transition proceeds by bubble nucleation, we
can study the propagation of bubbles in the hot plasma and the
transport properties of particles through the bubble wall. A
considerable amount of work has been devoted to this issue, and the
reader can consult the works of Ref. \cite{BW} for details. These
works assume that homogeneous nucleation theory is adequate to
investigate the evolution of the phase transition, at least for the
range of parameters of interest in the particular model being used to
generate the baryon asymmetry. This brings us to the second important
aspect of the nonequilibrium dynamics of first order phase
transitions, namely the validity of homogeneous nucleation theory to
describe the approach to equilibrium. This is the issue addressed in
the present work.

Nucleation theory is a well-studied, but far from exhausted, subject.
Since the pioneering work of Becker and D\"oring on the nucleation of
droplets in supercooled vapor \cite{BD}, the study of first order
phase transitions has been of interest to investigators in several
fields, from meteorology and materials science to quantum field theory
and cosmology. Phenomenological field theories were developed by Cahn
and Hilliard and by Langer \cite{CH,LANGER} in the context of
coarse-grained time-dependent Ginzburg-Landau models, in which an
expression for the decay rate per unit volume was obtained by assuming
a steady-state probability current flowing through the saddle-point of
the free-energy functional \cite{LANGER,DOMB}. The application of metastable
decay to quantum field theory was initiated by Voloshin, Kobzarev, and
Okun \cite{VKO}, and soon after put onto firmer theoretical ground by
Coleman and Callan \cite{CC}. The generalization of these results for
finite temperature field theory was first studied by Linde
\cite{LINDE}, and has been the focus of much recent attention
\cite{FINITETDECAY}.

The crucial ingredient in the evaluation of the decay rate is the
computation of the imaginary part of the free energy. As shown by
Langer \cite{LANGER}, the decay rate ${\cal R}$ is proportional to the
imaginary part of the free energy ${\cal F}$,
\beq
{\cal R} = {{\mid E_-\mid }\over {\pi T}}{\rm Im} {\cal F} ~,
\eeq
where $E_-$ is the negative eigenvalue related to metastability, which
depends on nonequilibrium aspects of the dynamics, such as the
coupling strength to the thermal bath. Since ${\cal F}= - T {\rm ln}
Z$, where $Z$ is the partition function, the computation for the rate
boils down to the evaluation of the partition function for the system
comprised of bubbles of the lower energy phase inside the metastable
phase. For a dilute gas of bubbles only, the partition function for
several bubbles is given by \cite{AM,LANGER},
\beqa
Z & \simeq & Z(\varphi_{f}) +Z(\varphi_{f}) \left[ \frac{Z(\varphi_{b})}
{Z(\varphi_{f})} \right] + Z( \varphi_{f}) \frac{1}{2 !} \left[
\frac{Z(\varphi_{b})}{Z(\varphi_{f})} \right]^{2} + \ldots
\nonumber \\
& \simeq & Z(\varphi_{f}) \exp \left[ \frac{Z(\varphi_{b})}{Z(\varphi_{f})}
\right]
\: ,
\label{e:Zmany}
\eeqa
where $\varphi_{f}$ is the metastable vacuum field configuration and
$\varphi_{b}$ is the bubble configuration, the bounce solution to the
$O(3)$-symmetric Euclidean equation of motion. We must evaluate the
partition functions above. This is done by the saddle-point method,
expanding the scalar field $\phi({\bf x},\tau)$,
such that $\phi({\bf x},\tau) \rightarrow \varphi_{f}
+\zeta(\bf{x} ,\tau)$ for $Z(\varphi_{f})$, and $ \phi({\bf x},\tau)
\rightarrow \varphi_{b} (\bf{x}) +\eta (\bf{x} ,\tau)$ for
$Z(\varphi_{b})$, where $\zeta(\bf{x} ,\tau)$ and $\eta (\bf{x} ,\tau)$
are small fluctuations about equilibrium.
Skipping details \cite{FINITETDECAY}, up to one-loop
order one obtains for the ratio of partition functions,
$\frac{Z(\varphi_{b})}{Z(\varphi_{f})}$,
\begin{equation}
\frac{Z(\varphi_{b})}{Z(\varphi_{f})} \stackrel{1-loop \: order}{\simeq}
\left[ \frac{\det ( -\Box_{E} + V''(\varphi_{b}))_{\beta}}
{ \det ( -\Box_{E} + V''(\varphi_{f}))_{\beta}} \right]^{-\frac{1}{2}}
e^{-\Delta S} \: ,
\label{ratiodet}
\end{equation}
where $[ \det (M)_{\beta}]^{- \frac{1}{2}} \equiv \int D \eta \exp
\left\{ - \int_{0}^{\beta} d \tau \int d^{3} x \frac{1}{2} \eta [M]
\eta \right\}$ and $\Delta S = S_{E}(\varphi_{b})-S_{E}(\varphi_{f})$
is the difference between the Euclidean actions for the field
configurations $\varphi_{b}$ and $\varphi_{f}$. [Note that
$S_{E}(\varphi)$, and hence $\Delta S$, does not include any
temperature corrections.] Thus, the free energy of the system is,
\beq\label{energy}
{\cal F} = - T \left[ \frac{ \det ( -\Box_{E} + V''(\varphi_{b}))_{\beta}}
{ \det ( -\Box_{E} + V''(\varphi_{f}))_{\beta}} \right]^{-\frac{1}{2}}
e^{-\Delta S} \: .
\eeq

We are briefly reproducing this computation here because we want to
stress the importance of the assumptions built into it. First, that
the partition function is given by equation \ref{e:Zmany} within the
dilute gas approximation, and second, that the partition function is
evaluated assuming {\it small} fluctuations about the homogeneous
metastable state $\varphi_{f}$. It is clear that for situations in
which there are large amplitude fluctuations about the metastable
equilibrium state the above formula must break down. Thus the
breakdown of the expression for the rate is intimately connected with
the question of how well-localized the system is about the metastable
state as the temperature drops below the critical temperature $T_c$.

This question has been addressed in the context of the electroweak
phase transition in the works listed in Ref. \cite{GK}. The common
assumption of these works is that for weak enough transitions the
dynamics is dominated by correlation-volume large-amplitude thermal
fluctuations, dubbed subcritical bubbles, which promote considerable
phase mixing as the Universe cools below the critical temperature.
Within the validity of this analytical approach it was shown that
homogeneous nucleation is only justified for Higgs masses below $70$
GeV or so, which is both dangerously close to the present lower bound
on the Higgs mass \cite{HIGGS}, and to the limit of validity of the
perturbatively evaluated effective potential \cite{VEFF}. Furthermore,
as with any analytical treatment of nonequilibrium dynamics, many
aspects of the complicated kinetics of the system are not included.
For example, even though correlation-volume bubbles may be the
dominant fluctuations about equilibrium, there will be bubbles of
different sizes present which may percolate and also acquire some
thermal velocity due to diffusive processes. It is clear that a final
answer to the problem can only be given by a combination of analytic
and numerical tools.

In order to clarify the distinction between `weak' and `strong'
first order transitions, one of us has recently studied the
nonequilibrium dynamics of a (2+1)-dimensional model of a scalar field
coupled to a thermal bath \cite{MG}. The nonlinear interactions were
chosen to reflect the gross properties of the electroweak effective
potential, although the model only deals with a real scalar field. It
was shown that there is a very clear distinction between a weak and a
strong transition, and that one should expect a very different
dynamics between the two cases. In the present work, we generalize
these results to (3+1)-dimensions and also address important issues
concerning the reliability of the numerical results. As we will show,
it is clear that the two regimes are easily distinguishable, as in the
(2+1)-dimensional case. Using homogeneous nucleation can easily lead
to the wrong description of the dynamics.

In passing, we note that somewhat similar results have been obtained
in the context of binary-fluid mixtures, where the (conserved) order
parameter is the local concentration of one of the components of the
mixture \cite{BARON}. If the system is quenched to concentrations
above the spinodal (the inflection point in the free-energy density),
the transition evolves by spinodal decomposition; otherwise,
nucleation occurs. The transition between the two regimes was shown in
that case to be smooth. However, we must remember that here we are
dealing with a non-conserved order parameter and have much faster
dynamics than in binary fluid mixtures. As we will show, the
transition between the two regimes is more dramatic in our context.

This paper is organized as follows. In the next Section we introduce
the model we will use in the numerical simulations. In Section 3 we
discuss details of the numerical approach used to study the dynamics
of the system, including the implementation of the code on a parallel
machine. In Section 4 we discuss the numerical results and their
reliability. We conclude in Section 5 with a summary of our results
and an outlook into future work.

\section {The Model}

The homogeneous part of the free energy density is written as
\beq
U(\p ,T)={a\over 2}\left (T^2-T_2^2\right )\p^2-
{{\alpha}\over 3}T\p^3 +{{\l}\over 4}
\p^4 \:.
\label{e:freen}
\eeq
This choice intentionally resembles the electroweak effective
potential to some order in perturbation theory, although here $\p({\bf
x},t)$ is a real scalar field, as opposed to the magnitude of the
Higgs field. The goal is to explore the possible dynamics of a model
described by the above free-energy density, generalizing the results
obtained in Ref. \cite{MG} to (3+1)-dimensions. The analogy with the
electroweak model is suggestive but not quantitative.

Introducing dimensionless variables ${\tilde x} = a^{{1}\over{2}} T_2
x$, ${\tilde t} = a^{{1}\over{2}} T_2 t$, $X = a^{-{{1}\over{4}}}
T_2^{-1} \p$, and $\theta = T/T_2$, the Hamiltonian is,
\beq
{{H[X]}\over {\theta}}={1\over {\theta}}\int d^2{\tilde x}\left [
{1\over 2}\mid {\tilde \bigtriangledown} X\mid^2 +
{1\over 2}\left (\theta^2 -1
\right )X^2 -{{
{\tilde \alpha}}\over 3}\theta X^3+{{{\tilde \l}}\over 4}X^4\right ] \:,
\label{e:hamilton}
\eeq
where ${\tilde \alpha} = a^{-{{3}\over{4}}} \alpha$, and ${\tilde
\lambda} = a^{-{{1}\over{2}}} \lambda$ (henceforth we drop the
tildes). For temperatures above $\theta_1=(1-\a^2/4\l)^{-{{1}\over
2}}$ there is only one minimum at $X=0$. At $\theta=\theta_1$ an
inflection point appears at $X_{\rm inf}=\alpha \theta_1 /2\lambda$.
Below $\theta_1$ the inflection point separates into a maximum and a
minimum given by $X_{\pm}={{\a\theta}\over {2\l}}\left [ 1\pm
\sqrt{1-4\l\left (1-1/\theta^2\right )/\a^2}\right ]$. At the critical
temperature $\theta_c=(1-2\a^2/9\l)^{-{{1}\over 2}}$ the two minima,
at $X_{\rm o}=0$ and $X_+$ are degenerate. Below $\theta_c$ the minimum at
$X_+$ becomes the global minimum and the $X_{\rm o}$-phase becomes
metastable. Finally, at $\theta=1$ the barrier between the two phases
at $X_-$ disappears.

The coupling with the thermal bath will be modelled by a Markovian
Langevin equation which, in terms of the dimensionless variables
defined above, is
\beq\label{e:langevin}
{\partial^2X\over\partial t^2} = \nabla^2 X - \eta {\partial X\over
\partial t} - {\partial U(X,\theta) \over \partial X} + \xi(x,t)~~,
\eeq
where $\eta$ is the dimensionless viscosity coefficient, and $\xi$ the
dimensionless stochastic noise with vanishing mean, related to $\eta$
by the fluctuation-dissipation theorem,
\beq\label{e:fluc-diss}
\langle \xi({\bf x},t)\xi( {\bf x'},t')\rangle =
  2 \eta \theta \delta(t-t')\delta^3({\bf x} - {\bf x'})~~.
\eeq
A few comments are in order concerning our choice of equation. It is
clear that we are assuming that $X({\bf x},t)$ represents the
long-wavelength modes of the scalar field. Whenever one discretizes a
continuum system there is an implicit coarse-graining scale built in.
We encapsulate information about the shorter-wavelength modes, which
have faster relaxation time-scales, in the dissipation and noise
terms. In principle it should be possible to derive an effective
Langevin-like equation for the slow modes by integrating out the fast
modes from the effective action. This is a complicated problem, and
progress has been slow. Recent work indicates that one should expect
departures from the Langevin equation written above \cite{LANGEVIN},
although details are sensitive to the particular model one starts
with. For example, the noise may be colored (with more complicated
correlation functions) and the coupling to the bath may be
multiplicative, as opposed to the additive coupling chosen above.
Here, we will adopt the above equation as a first step. We do not
expect that the nature of the noise will change the final equilibrium
properties of the system, but mostly the relevant relaxation
time-scales. Since the physical results here are related to the final
equilibrium state of the system, we believe that they will not be
affected by more complicated representations of the coupling of the
field to the thermal bath. However, a more thorough examination of this
question deserves further study.

A related topic is the choice of coarse-graining scale, which is
embedded in the lattice spacing used in the simulations. It is
well-known that any classical field theory in more than one spatial
dimension is ultra-violet divergent, and that the lattice spacing
serves as an ultra-violet cutoff. This being the case, one should be
careful when mapping from the lattice to the continuum theory. If one
is to probe physics at shorter wavelengths, renormalization
counterterms should be included in the lattice formulation so that a
proper continuum limit is obtained on the lattice within the validity of
perturbation theory. This point has been
emphasized in Ref. \cite{AG}, where a (2+1)-dimensional study of
nucleation was performed for a temperature-independent potential.
Renormalization counterterms (of order $\theta ~{\rm ln}\,\delta x$
for lattice spacing $\delta x$) for a particular renormalization
prescription were obtained, and the results shown to be lattice-space
independent.

Here, due to the temperature dependence of the potential, the
renormalization prescription of Ref. \cite{AG} does not work.
Instead, we will use $\delta x = 1$ throughout this work. It turns out
that for all cases studied the mean-field correlation length
$\xi^{-2}=V^{\prime \prime} (X_{\rm o},\theta_c)$ is sufficiently larger
than unity to justify this choice. Modes with shorter wavelengths
are coupled through the noise into the dynamics of the longer
wavelength modes, as described by equation \ref{e:langevin} above.

\section{Lattice Formulation}

The system is now discretized onto a lattice of length $L$ with grid
spacing $\delta x$, time step $\delta t$, and total run time $\Delta
t$. Using a standard second order staggered leapfrog approach equation
\ref{e:langevin} becomes
\beqa
\dot{X}_{i,n+1/2} & = & \frac{1}{1 + \frac{1}{2}\eta t_n}
\left[ \left(1-\frac{1}{2}\eta t_n \right)\dot{X}_{i,n-1/2}
+ \delta t \left( \nabla^2 X_{i,n}
- \left. \frac{\partial U}{\partial X} \right|_{i,n}
+ \xi_{i,n} \right) \right] \nonumber \\
X_{i,n+1} & = & X_{i,n} + \delta t \; \dot{X}_{i,n+1/2}
\eeqa
where $i$-indices are spatial and $n$-indices temporal. The continuum
white noise $\xi({\bf x}, t)$ is replaced by its discretized analogue
$\xi_{i,n}$ by requiring that the discrete noise be uncorrelated on
all scales above the shortest simulated. The discretised form of the
fluctuation-dissipation relation of equation \ref{e:fluc-diss} becomes
\beq
\langle \xi_{i_1,n_1} \xi_{i_2,n_2} \rangle = 2 \eta \theta \,
\frac{1}{\delta t}\,\delta_{{n_1},{n_2}}\,
\frac{1}{\left (\delta x\right )^3}\,\delta_{{i_1},{i_2}}~~.
\eeq
The discrete noise is hence approximated by
\beq
\xi_{i,n} = \sqrt{\frac{2 \eta \theta}{\delta t \,\left (\delta x\right )^3}}
\;{\cal G}_{i,n}
\eeq
where ${\cal G}_{i,n}$ is a unit-variance Gaussian random number at
each point in lattice spacetime.

Since we are modelling an unbounded system we do not want our
simulation volume to have a distinct surface; we therefore use
periodic boundary conditions. However, such boundary conditions may
induce errors if a simulation runs for longer than a time causally
equivalent to $L/2$ as the periodicity then introduces spurious
long-range correlations. Since we must run very long simulations to
guarantee equilibration, were this constraint to apply we would be
forced to use impossibly large lattices. Fortunately, the presence of
the noise term, uncorrelated at each point in lattice spacetime, has
the effect of swamping any such effect.

We must now run our simulations on large lattices (to reduce finite
size effects) many times over (to reduce statistical noise) and for
long run times (to ensure reaching equilibrium). Typically any attempt
to reduce one of these constraints is met by a corresponding increase
in another --- for example, smaller grids give noisier results
requiring more runs. As a consequence, we soon find ourselves at the
limits of what is possible on a workstation. We have therefore
parallelised the code and implemented it on a 128-node AP1000. The
overall lattice is subdivided into an appropriate number of
sub-lattices, each of which is local to a single node. These
sub-lattices are defined with an overlap, such that each edge of any
node's sub-lattice is included within the body of one of its
neighbours. At each time step each node evolves the body of its local
sub-lattice, but not the edges, for which insufficient data is locally
available to calculate spatial derivatives. Each node then swaps the
necessary data to update the overlapping edges of their associated
sub-lattices with each of its neighbouring nodes.

The one qualitatively different feature of the parallel code is in the
implementation of the random number generator. Computational random
number generators are required to produce the same sequence of numbers
whenever they are given the same initial conditions. Therefore what
they actually generate, at least ideally, is a predictable, periodic,
sequence of pseudo-random numbers, any sufficiently short ({\it i.e.}
substantially shorter than the period) subset of which has statistical
properties indistinguishable from a genuine random sequence. In our
case we need random numbers at every point in lattice spacetime which
are uncorrelated across the entire simulation spacetime; each node
requires numbers which are random not only locally at the node itself,
but also across all the other nodes too. Either each node must have a
different generator (highly impractical for any more than a few nodes)
or each node must be allocated a unique subsequence of a single
generator's full sequence. Most generators are based on an iterative
scheme, where each number in the sequence is calculated from some of
the previous numbers in the sequence. However, if these previous
numbers are not members of the local subsequence then there is a
communication cost incurred in fetching them from the relevant node.
In order to maximise the efficiency of the code, we require a
generator whose sequence can be broken down into subsequences with
elements generated by reference to previous members of that
subsequence alone. Moreover, for large lattices and long run times we
require a generator with an extremely long period. Thus for $L = 48$,
$\delta t = 0.1$, and a running to $\Delta t = 3000$ (values used
below) we require of the order of $2^{31}$ random numbers, and hence a
generator with a period many times longer than this. A solution to
this problem, applicable across any $2^n$ nodes, with a period of
$2^{607}-1$ and with each element in each subsequence calculable
completely locally, is given by a particularly elegant parallelisation
of the generalised feedback shift register algorithm \cite{RNG}; this
is the generator implemented here.

\section{Numerical Experiment and Results}

As pointed out in the Introduction, the question as to whether a first
order phase transition is `weak' or `strong' boils down to how well
localized in the metastable state the system is as the temperature
drops below the critical temperature. In order to address this
question, following the procedure of Ref. \cite{MG}, we will study the
behavior of the system at the critical temperature, when the two
minima are degenerate. The reason for this choice follows naturally
from the fact that we are interested on the way by which the system
approaches equilibrium as the temperature drops below $T_c$. The
detailed dynamics will depend on the relative fraction of the total
volume occupied by each phase; if at $T_c$ the system is well
localized about the $X=0$ minimum, as the temperature drops the
transition may evolve by nucleation and subsequent percolation of
bubbles larger than a critical size.  If, on the other hand,
considerable phase-mixing occurs already at $T_c$, we expect the
transition to evolve by domain coarsening, with the domains of the
$X_+$ phase eventually permeating the whole volume.

Let us call the two phases the $0$-phase and the $+$-phase,
corresponding to the local equilibrium values $X=X_{\rm o}=0$, and $X=X_+$,
respectively. We can quantify the phase distribution of the system as
it evolves according to equation \ref{e:langevin}, by measuring the
fraction of the total volume in each phase. This is done by simply
counting the total volume of the system at the left of the potential barrier's
maximum
height ({\it i.e.}, $X\leq X_-\equiv X_{\rm max}$),
corresponding to the $0$-phase.
Dividing by the total volume, we obtain the fraction of the system in
the $0$-phase, $f_{\rm o}(t)$, such that
\beq
f_{\rm o}(t) + f_+(t) = 1 ~~,
\eeq
where, of course, $f_+(t)$ corresponds to the fractional volume in the
$+$-phase.

A further measure of any configuration is given by the volume-averaged
order parameter, $\langle X \rangle (t)= V^{-1}\int dV~X(t)$. A
localised configuration ($f_{\rm o}^{\rm eq} > 0.5$) then corresponds
to $\langle X \rangle^{\rm eq} < X_{\rm max}$, and a fully phase-mixed
configuration ($f_{\rm o}^{\rm eq} \simeq 0.5$) to $\langle X
\rangle^{\rm eq} = X_{\rm max}$, where the superscript `eq' refers
to final ensemble-averaged equilibrium values of $f_{\rm o}(t)$ and
$\langle X \rangle (t)$.

We prepare the system so that initially it is well localised in the
$0$-phase, with $f_{\rm o}(0)=1$ and $\langle X \rangle (0)=0$. These
initial conditions are clearly the most natural choice for the problem
at hand. If one has cosmology in mind, it is quite possible that as
the system slowly cools down (we are not interested in phase
transitions close to the Planck scale), fluctuations from the high
temperature phase $X=0$ to the $X_+$ phase are already occuring before
$T_c$ is reached. (In this case, our arguments are even stronger.)
However, we will adopt the best-case scenario for homogeneous
nucleation to work, in which the system managed to reach the $X=0$
phase homogeneously, so that the initial state is a thermal state with
mean at $X_{\rm o}$.  If one has more concrete applications in mind, we can
assume that we quenched the system to its critical temperature, making
sure that the order parameter remains localized about the
high-temperature phase.  Since thermalization happens very fast in the
simulations, the exact point by point initial conditions should not be
important, and we can view the first few time steps as generating an
initial thermal distribution with $f_{\rm o}(0) \sim 1$ and $\langle X
\rangle \sim 0$, so that the average kinetic energy per lattice point
satisfies the equipartition theorem, ${1\over N}E_k = {3\over 2}T$.
For simplicity we take $X=0, \dot X=0$ everywhere initially.

There are two parameters controlling the strength of the transition,
$\alpha$ and $\lambda$. In the previous (2+1)-dimensional work,
$\alpha$ was chosen to vary while $\lambda$ was kept fixed. It is
really immaterial which parameter is held fixed, or if both are made
to vary, but in order to keep closer to the spirit of the electroweak
model we will fix $\alpha$ and let $\lambda$ vary. As is well-known,
$\lambda$ is related to the Higgs mass, while $\alpha$ is related to
the gauge-boson masses \cite{EW}. The connection with the
electroweak model is straightforward. If we consider as an example the
unimproved one-loop approximation, the effective potential is
\cite{EW},
\beq
\label{eq:VEW}
V_{{\rm EW}}(\p,T)=D\left (T^2-T_2^2 \right )\p^2-ET\p^3+{1\over
4}\l_T\p^4,
\eeq
where $D$ and $E$ are given by
$D=\left[6(M_W/\s)^2+3(M_Z/\s)^2+6(M_T/\s)^2\right ]/24\simeq 0.17$
and $E=\left [6(M_W/\s)^3+3(M_Z/\s)^3\right ]/12\pi\simeq 0.097$, for
$M_W=80.6$ GeV, $M_Z=91.2$ GeV, $M_T=174$ GeV \cite{MT}, and $\s=246$
GeV. $T_2$ is given by,
\beq
\label{eq:T2}
T_2=\sqrt{(M_H^2-8B\s^2)/4D}\ ,
\eeq
where the physical Higgs mass is given in terms of the 1-loop
corrected $\l$ as $M_H^2=\left (2\l+12B\right) \s^2$, with $B=\left
(6M_W^4+3M_Z^4-12M_T^4\right )/64\pi^2\s^4$, and the
temperature-corrected Higgs self-coupling is,
\beq
\l_T=\l-{1\over {16\pi^2}}\left [
\sum_Bg_B\left ({{M_B}\over {\s}}\right )^4
{\rm ln}\left (M_B^2/c_BT^2\right )-\sum_Fg_F\left ({{M_F}\over {\s}}
\right )^4{\rm ln}\left (M_F^2/c_FT^2\right )\right]
\eeq
where the sum is performed over bosons and fermions (in our case only
the top quark) with their respective degrees of freedom $g_{B(F)}$,
and ${\rm ln}c_B=5.41$ and ${\rm ln}c_F=2.64$.

Thus, the correspondence with our (dimensionless) parameters is
\beq
\alpha = {{3E}\over {(2D)^{3\over 4}}}=0.065~,~{\rm and}~~~\l = {{\l_T}\over
{(2D)^{1\over 2}}}=1.72 \l_T~~.
\eeq
Once this is established, the numerical experiment proceeds as
follows: i) Choose $\alpha=0.065$; ii) Prepare the system in the initial
state described above, and measure the value of $f_{\rm o}(t)$ and
$\langle X\rangle (t)$ for several values of $\lambda$, as the system
evolves according to equation \ref{e:langevin}. As with any numerical
experiment, one must make sure that the results are independent of
lattice artifacts (or at least the dependence is understood), such as
its finite size and choice of time-step for the evolution routine.
Before we discuss our results, we will address these issues in the
following Subsection.

\subsection{Finite Lattices and the Thermodynamic Limit}

Whenever simulating a system on the lattice, we are faced with the
hard decision of having to achieve a compromise between approximating
the infinite volume limit, and having fairly reasonable integration
times. This problem is particularly serious in the context of phase
transitions, as it is well-known that symmetry breaking only occurs in
the infinite volume limit; at finite volumes, there is a nonzero
probability that a large fluctuation will restore the broken symmetry.
Even though this is formally true, we will argue here that this does
not represent a problem to our simulations, if we make sure that the
lattice is large enough. There is a large amount of literature on
finite-size effects and how they are handled in different contexts
\cite{FINITESIZE}, and we do not intend to reproduce these results.
What we want to do is to bring this issue closer to our problem.

We are interested in studying the system given by the free-energy
density of equation \ref{e:freen}, at the critical temperature
$\theta_c$ when the two minima are degenerate. The system is prepared
in the $0$-phase, and we measure the fraction of the volume in each
phase as it evolves. We will give a rough estimate of how large the
lattice should be in order to suppress spurious symmetry restoration
(that is, $f_{\rm o}\rightarrow 0.5$, in our case) due to the lattice size.
There are two relevant time-scales in the problem, the relaxation
time-scale for small-amplitude fluctuations within the $X=0$ well,
$\tau_{\rm rel}$, and the `escape' time-scales for large amplitude
fluctuations into the $+$-phase, $\tau_{\rm esc}$. In terms of the
rate per unit volume for each process, we write the relevant
time-scales as
\beq
V\Gamma_{\rm rel}\sim \tau_{\rm rel}^{-1} \sim T\gamma_{\rm rel}^{-1}~~,
\eeq
where $\gamma_{\rm rel}$ is the typical relaxation time-scale for
short amplitude fluctuations in units of $T^{-1}$, and
\beq
V\Gamma_{\rm esc} \sim \tau_{\rm esc}^{-1} \sim T{\rm exp}[-F_f/T]~~,
\eeq
where $F_f$ is the free energy of the large amplitude fluctuation. The
condition for large amplitude fluctuations to be suppressed in
comparison to typical relaxation processes is then,
\beq
{{ \tau_{\rm esc}}\over {\tau_{\rm rel}}}\gg 1 ~~~~\Rightarrow ~~~~~~
{\rm exp}[F_f/T]\gg \gamma_{\rm rel}~~.
\label{e:finitevol}
\eeq
To estimate $\tau_{\rm esc}$ note that within the Gaussian
approximation, a homogeneous fluctuation of volume $V$ and amplitude
$\phi_A$ about equilibrium ($\phi=0$, for simplicity) has free energy
\beq
F_f(\phi_A,V,T)={V\over 2}m^2(T)\phi_A^2~~,
\eeq
where $m^2(T)=V''(\phi=0,T)$ (we neglect the gradient contribution, as
it would suppress the fluctuation even further, making our arguments
stronger). We are interested in fluctuations about $X=0$ (we now go
back to our dimensionless variables), at the critical temperature
$\theta_c = (1-2\alpha^2/9\lambda)^{-{{1}\over 2}}$. We expect growing
instabilities to be triggered whenever fluctuations probe the
nonlinearities in the free energy. This is corroborated by the results
in Ref. \cite{MG} and, as we will soon see, here also. Thus we
consider fluctuations with amplitude equal to the nearest inflection
point to $X=0$, namely $X_A={{\alpha\theta_c}\over{3\lambda}}\left
(1-1/\sqrt{3}\right)$. Writing their volume $V = {{4\pi}\over 3}\left
(n\xi \right )^3$ in terms of the correlation length $\xi
(\theta_c)=(\theta^2_c -1)^{-{{1}\over 2}}$, with $n$ a real number,
we obtain
\beq
{{F_f}\over {\theta_c}}\simeq 0.088{{\alpha}\over {\lambda^{3/2}}}n^3~~.
\eeq
{}From our arguments above it is clear that $\gamma_{\rm rel} = \left (
\theta^2_c-1\right )^{-{{1}\over 2}}$. Let us consider an example which
will be relevant later on. Take $\alpha=0.065$ and $\lambda=0.020$. In
this case, the correlation length is $\xi(\theta_c)\simeq 4.5$ and we
obtain, from equation \ref{e:finitevol}, $n\gg 0.91$. Since the radius
of the fluctuation is $R_f=n\xi$ this result implies that fluctuations
probing the inflection point with radius $R_f= 0.91\times 4.5\simeq
4.1$ are probable within typical relaxation time scales of the system.
Thus the lattice length $L$ should be sufficiently larger than about
$2 R_f$ ($L \geq 10$ or so) to avoid spurious symmetry restoration.
For $L=20$ the volume ratio of the lattice to the above fluctuations
is around $L^3/4R_f^3\sim 25$, and such processes alone cannot restore
the symmetry within time-scales of interest in the dynamics. In all
the results quoted in this work we use $L=48$.

We give two pieces of evidence supporting these arguments. In Fig.1
the equilibrium values of the $0$-phase fraction, $f_{\rm o}^{\rm eq}$, and
of the volume-averaged order parameter, $\langle X\rangle^{\rm eq}$,
are given as a function of the lattice size $L$, for $\alpha=0.065$
and $\lambda=0.020$. $\lambda$ is chosen so that for large lattices
the system remains localized mostly in the $0$-phase. Note that as $L$
decreases $ f_{\rm o}^{\rm eq}\rightarrow 0.5$, representing a spurious
`symmetry restoration' due to the smallness of the lattice. For small
lattices, the two phases are completely mixed. In fact, changes can be
seen between $L=10$ and $L=20$, in qualitative agreement with our
arguments above. Note that for large enough lattices, the equilibrium
value approaches a stable value which is independent of the lattice
size. For all practical purposes, this is the infinite volume limit.
Fluctuations large enough to restore the symmetry are possible, but
with negligible probability.

The reader may wonder why for large enough lattices $f_{\rm o}^{\rm eq}\neq
1$. The reason for this is that at finite temperatures there is a
nonvanishing probability per unit volume of having thermal
fluctuations populating the $+$-phase. Even though these fluctuations
are unstable and shrink away, there will be an equilibrium
distribution of bubbles suppressed by a Boltzmann factor. For very
strong transitions (very small $\l$), $f_{\rm o}^{\rm eq}
\sim 1$, and a negligible fraction of the system is in the $+$-phase.
We refer the reader to the paper by Gelmini and Gleiser, Ref.
\cite{GK} for details. In Fig. 2 we show a phase space portrait of the
system for a given point on the lattice for different lattice sizes.
That is, we choose a particular point $X(x_{\rm o},y_{\rm o},z_{\rm o},t)$ and
follow
its evolution, making a plot of ${\dot X}(x_{\rm o},y_{\rm o},z_{\rm o},t)$ vs
$X(x_{\rm o},y_{\rm o},z_{\rm o},t)$. Taking $\alpha=0.065$ and
$\lambda=0.010$, Fig. 2a
shows the $L=4$ and Fig. 2b the $L=20$ case. It is clear that for the
smaller lattice the system is probing both minima of the free energy
($X_{\rm o}=0$ and $X_+=4.55$ here), while for the larger lattice the system
remains localized in the $X=X_{\rm o}$ well. As the lattice size is
increased the throat separating the two minima becomes less and less
dense until eventually the system becomes `trapped' within one well.
The time-scale for its eventual escape is much larger than any
time-scale of interest in this problem, with $\tau_{\rm esc}
\rightarrow \infty$ as $L\rightarrow \infty$.

Finally, in Fig. 3, we show the equilibrium values of the fraction
$f_{\rm o}(t)$, $f_{\rm o}^{\rm eq}$, and of the volume-averaged order
parameter,
$\langle X\rangle^{\rm eq}$, for $\alpha=0.065$ and $\lambda=0.020$,
as a function of time step, $\delta t$. Note that using too large a
time step compromises the stability of the simulations, tending to
drive the system's equilibrium towards the phase-mixed symmetric state
($f_{\rm o}^{\rm eq} \rightarrow 0.5$, $\langle X\rangle^{\rm eq} = X_{\rm
max}$).  The results detailed below are therefore obtained using a
time step $\delta t = 0.1$.

\subsection{Numerical Results}

Based on the above discussion, we choose $L=48$, $\delta x = 1$,
$\delta t = 0.1$, and $\alpha = 0.065$ in all simulations. The
experiment then consists in measuring the fraction of the volume in
the $0$-phase as a function of time for several values of $\lambda$.
As this involves a stochastic noise, we must perform an ensemble
average over many runs in order to obtain physically reasonable
results.

In Fig. 4 we show the evolution of the ensemble-averaged fraction
$f_{\rm o}(t)$ for several values of $\l$. It is clear that for small enough
values of $\lambda$ the system remains well-localized in the $0$-phase
with $f_{\rm o}^{\rm eq}\sim 1$, while for larger values the two phases
become completely mixed, with $f_{\rm o}^{\rm eq}\rightarrow 0.5$.
Remarkably, the transition region between the two regimes is quite
narrow, centered around $\l \simeq 0.025$. This can be seen from Fig.
5 where we show $f_{\rm o}(t)$ for $\l = 0.024,~0.025$, and $0.026$. [The
curves are noisier due to the fact that we must run for longer times in
order to approach the equilibrium values of $f_{\rm o}(t)$, being thus
constrained to perform an ensemble average with fewer runs.] Note that
for $\l=0.026$, $f_{\rm o}^{\rm eq}\simeq 0.5$, while for $\l = 0.024$,
$f_{\rm o}^{\rm eq} \simeq 0.72$. There is a pronounced change in the
behavior of the system for $\l \simeq 0.025$. Furthermore, we find
that the numerical curves can be fitted at all times by a stretched
exponential,
\beq
\label{e:expfit}
f_{\rm o}(t) = \left (1-f_{\rm o}^{\rm eq}\right ){\rm exp}
\left [-(t/\tau_{\rm eq})^{\s}\right ] + f_{\rm o}^{\rm eq}~~,
\eeq
where $f_{\rm o}^{\rm eq}$ is the final equilibrium fraction and
$\tau_{\rm eq}$ is the equilibration time-scale. In Table 1 we list
$\s$ and $\tau_{\rm eq}$ for several values of $\lambda$. Note that
for $\l = 0.025$ the fit is obtained at late times by a power law,
(smooth dashed curve in Fig. 5)
\beq
\label{e:powerlaw}
f_{\rm o}(t)\mid_{\l=0.025}~~  \propto ~~t^{-k}~~,
\eeq
with $k = 0.10(\pm 0.02)$. This slowing down of the approach to
equilibrium is typical of systems in the neighborhood of a second
order phase transition, being known as `critical slowing down'
\cite{SLOW}.  This behavior is suggestive of a phase
transition between two possible regimes for the system, one in which
the system is well-localized in the $0$-phase, and the other in which
there is a complete mixing between the two phases.  Let us call these
two regimes the `strong' and the `weak' regimes, respectively.  Before
we explore this idea any further, it is prudent to check if the final
equilibrium fractions are sensitive to the viscosity parameter $\eta$,
which reflects the coupling of the system to the thermal bath. In Fig.
6 we show the approach to equilibrium for several values of $\eta$ and
for $\l=0.020$, which lies within the `strong' regime. It is clear
that whereas the equilibration time-scales are sensitive to the value
of $\eta$, with larger $\eta$ implying slower equilibration, the {\it
final} equilibrium fractions are the same. This is precisely what one
expects, as the coupling to the bath should not influence the final
equilibrium properties of the system, but only how fast it
equilibrates.

Armed with these results, and invoking also the results in $(2+1)$-dimensions
\cite{MG}, we define the equilibrium fractional population difference
\beq
 \Delta F^{\rm eq}\left (\theta_c \right ) = f_{\rm o}^{\rm eq} - f_+^{\rm
eq}~~.
\eeq
In Fig. 7 we show the behavior of $\Delta F^{\rm eq}$ as a function of
$\l$.  There is a clear qualitative analogy between the behavior of
$\Delta F^{\rm eq}$ as a function of $\l$ and the behavior of the
magnetization as a function of temperature in Ising models.  Here, the
order parameter is the equilibrium fractional population difference
and the control parameter is $\l$. $\l_c\simeq 0.025$ is the critical
value for the parameter $\l$, which determines the degree of mixing
of the system at $T_c$.

We stress that the idea here is to probe the assumption of
localization within the $0$-phase as the system cools to $T_c$. Our
results show that if the time-scales for cooling are slower than the
equilibration time-scales of the system, for $\l > \l_c$ there will be
considerable phase mixing before the temperature drops below $T_c$.
This result can be made quite transparent by comparing the equilibrium
value of the volume-averaged field $\langle X\rangle^{\rm eq}$, and
the location of both the inflection point and the maximum of the
potential with varying $\l$. As can be seen from Fig. 8, the narrow
transition region is clearly delimited by
\beq
X_{\rm inf}~~ <~~ \langle X\rangle^{\rm eq} ~~< X_{\rm max}~~,
\eeq
where $X_{\rm inf}$ and $X_{\rm max}$ are the inflection point and the
maximum of the potential barrier, respectively. Note that for $\l \geq
0.026$, $f_{\rm o}^{\rm eq} = 0.5$ and $\langle X\rangle^{\rm eq} = X_{\rm
max}$. Recalling the information from Fig. 7, we conclude that there
is a clear distinction between the `strong' and `weak' regimes. The
existence of a potential barrier between the two phases at $T_c$ does
not imply in the system being localized in the $0$-phase, for large
enough values of $\l$. This efficient thermal mixing of the phases
will affect the dynamics of the phase transition as the temperature
drops below $T_c$.  For $\l > 0.025$, if the cooling is slow enough,
there is no reason to expect that the system will approach its final
equilibrium by nucleation of bubbles larger than critical. Rather, the
mechanism will resemble spinodal decomposition, where domains of the
two phases will compete for dominance, with the $+$-phase eventually
dominating the volume. We stress that the value $\l_c = 0.025$ is a
weak lower bound; even if the system starts localized in the $0$-phase
at $T_c$ for a smaller value of $\lambda$, the potential barrier will
also decrease for lower temperatures, and we should expect departures
from nucleation settling in even for $\l < 0.025$. [Of course, the
decrease in temperature will also suppress thermal fluctuations.] The
evaluation of the exact value of $\l$ for which nucleation theory will
break down depends on the asymmetry of the potential, the temperature,
and on the cooling rate, requiring further study. However, we can
assert that this value will be at least lower than $\l_c$.

\section{Conclusions and Outlook}

In this paper we investigated the possibility that thermal
fluctuations may induce considerable phase mixing for systems which
exhibit two degenerate phases at their critical temperature. By
modelling the nonequilibrium dynamics by means of a Langevin equation
with the system initially localized in one phase, we showed that
complete mixing of the two phases can occur, despite the presence of a
potential barrier between the phases. These results are of importance
in the context of cosmological phase transitions, in particular when
the cooling rate is sufficiently slow compared to the equilibration
time-scales of the system. This limit is implicit in our simulations,
as we held the temperature fixed at $T_c$ while the system
equilibrated. Our results should also be of importance for systems
studied in the laboratory. As we mentioned in the Introduction, binary
fluid mixtures separate by spinodal decomposition if the system is initially
quenched with concentrations above the spinodal. The dynamics in that
case is much slower though, and the transition between nucleation and
spinodal decomposition is smoother \cite{BARON}. However, we believe
the essential physics to be the same, as we found that considerable
phase mixing occured precisely as the equilibrium value of the
volume-averaged scalar field (the equivalent of the concentration in
the binary fluid case) traversed within the `transitional' region
delimited by the locations of the inflection point and the maximum of
the potential barrier. Also, as mentioned in previous publications
\cite{GK}, this phase mixing is characteristic of pre-transitional
phenomena found in the isotropic-nematic transition of liquid
crystals, which is known to be `weakly' first order \cite{LIQUID}.

Inspired by the effective potential of the electroweak model, we
measured the degree of mixing with respect to the value of the quartic
self-coupling of the scalar order parameter. However, as mentioned
before, this is not a simulation of the electroweak transition, as our
order parameter is a real scalar field. In fact, the critical value
above which we found that the two phases mix, $\l_c \simeq 0.025$, is
below the value of $\l$ which corresponds to a physical Higgs mass.
Using the equivalence relation between the two models obtained in
equation 16, a Higgs mass of $M_H = 50$ GeV would correspond to $\l =
0.0518$, which is well within the `weak' regime. For this value of
$\l$, the equilibrium value of the volume averaged field $\langle X
\rangle^{\rm eq} = X_{\rm max}$, and the two phases are
indistinguishable.

The results presented here are in qualitative agreement with a
previous work in which phase mixing was investigated in
(2+1)-dimensions \cite{MG}. As in that case, we suggest that the
distinction between `weak' and `strong' first order transitions be
quantified in terms of the equilibrium fractional population difference between
the two phases, $\Delta F^{\rm eq} = f_{\rm o}^{\rm eq} - f_+^{\rm eq}$.
Within the limits of a finite lattice, we observed a sharp
change in the behavior of $\Delta F^{\rm eq}$ with respect to
$\l$, (see Fig. 7) which we suggest characterizes a second order phase
transition between the `symmetric' (`weak') phase $\Delta F^{\rm eq} =
0$ and the `broken-symmetric' (`strong') phase $\Delta F^{\rm eq} =
1$. In order to study the details of the transition, a more thorough
analysis of finite-size effects in the determination of the critical
value $\l_c\simeq 0.025$ and of the critical exponent $\beta$,
obtained from the relation $\Delta F^{\rm eq} = (\l_c - \l)^{\beta}$
in the neighborhood of $\l_c$, must be performed. But taken together,
the sharp change in $\Delta F^{\rm eq}$ and the presence of
critical slowing down about $\l_c$ provide substantial evidence for
the existence of this transition.

Finally, our work also calls for a more detailed analysis of the
r\^ole of noise with spatio-temporal memory in simulations of the
dynamics of phase transitions in field theories. The Markovian
Langevin equation used here is certainly an approximation to more
complicated couplings between the system and the thermal bath. It
remains to be seen what r\^ole will colored noise play in the
nonequilibrium dynamics of field theories. It is however hard to
imagine that the nature of the noise can affect final equilibrium
properties of the system, although it may affect the equilibration
time-scales. In this connection, we note that recent results by one of
us show that at high enough temperatures, colored noise becomes white
\cite{LANGEVIN}. Given our present level of understanding of the
nonequilibrium dynamics of field theories it is only fair to expect
many surprises in the coming years.

\acknowledgments

JB acknowledges the support and assistance of the Imperial College /
Fujitsu Parallel Computing Research Center. MG was supported in part
by a National Science Foundation Presidential Faculty Fellows Award No.
PHY-9453431
and by grant No. PHYS-9204726. This collaboration was made possible by
NATO Collaborative Research Grant `Cosmological Phase Transitions' no.
SA-5-2-05 (CRG 930904) 1082/93/JARC-501.

\listoffigures

\noindent Figure 1.  The equilibrium values of the $0$-phase
fraction $f_{\rm o}^{\rm eq}$ and of the volume-averaged
field $\langle X \rangle^{\rm eq}$ are shown as a function of lattice length
$L$ for $\l=0.020$.\\

\noindent Figure 2a. Phase space
portrait for an arbitrarily chosen point of a lattice
of length $L=4$. $\l=0.010$ in this run. Note how the system probes both minima
of the potential. \\

\noindent Figure 2b. Phase
space portrait for an arbitrarily chosen point of a lattice
of length $L=20$. $\l=0.010$ in this run. Note how the system remains
constrained within the $0$-phase.\\

\noindent Figure 3.  The equilibrium values of the $0$-phase
fraction $f_{\rm o}^{\rm eq}$ and of the volume-averaged
field $\langle X \rangle^{\rm eq}$ are shown as a function of time step
$\delta t$ for  $\l=0.020$.\\

\noindent Figure 4.  Evolution of the ensemble-averaged
fraction $f_{\rm o}(t)$ for several values
of $\l$. \\

\noindent Figure 5.   Evolution of the ensemble-averaged
fraction $f_{\rm o}(t)$ for a narrow range about the critical value
$\l=0.025$. The smooth dashed curve is the power law fit described in
equation \ref{e:powerlaw}.\\

\noindent Figure 6.   Evolution of the ensemble-averaged
fraction $f_{\rm o}(t)$ for several values
of the viscosity parameter $\eta$ and for $\l=0.020$. Note how the final
equilibrium fraction is independent of the value of $\eta$.\\

\noindent Figure 7.  The fractional equilibrium
population difference $\Delta F^{\rm eq}$ for several values of $\l$. Note
the sharp discontinuity about the critical value $\l_c\simeq 0.025$.\\

\noindent Figure 8.  Comparison of the equilibrium value of the
volume-averaged field $\langle X \rangle^{\rm eq}$ (circles), with the
location of the nearest inflection point to $X=0$, $X_{\rm inf}$, and
the potential barrier, $X_{\rm max}$, as a function of $\l$. Note the
existence of a transitional region for $0.021 \simeq \l < 0.026$.  For
$\l \geq 0.026$, $\langle X \rangle^{\rm eq} = X_{\rm max}$.\\

\listoftables

\noindent Table 1.  The values of the equilibration time-scales and
the exponents for the exponential fit of equation \ref{e:expfit} for
several values of $\l$.\\

\vfill\eject

\centerline {\bf TABLE 1}
\vspace{2.in}

\begin{center}
\begin{tabular}{|c|c|c|}\hline\hline
$\l $ & $\tau_{{\rm EQ}}$ & $\sigma $ \\
\hline\hline
 0.015 & 25.0  & 1.0 \\
 0.020 & 45.0  & 1.0 \\
 0.022 & 60.0  & 1.0  \\
 0.024 & 110.0 & 0.80   \\
 0.026 & 220.0 & 0.50 \\
 0.028 & 120.0 & 0.75 \\
 0.030 & 100.0 & 0.80 \\
 0.035 & 55.0  & 0.80  \\
 0.040 & 40.0  & 0.90 \\
\hline
\end{tabular}
\end{center}

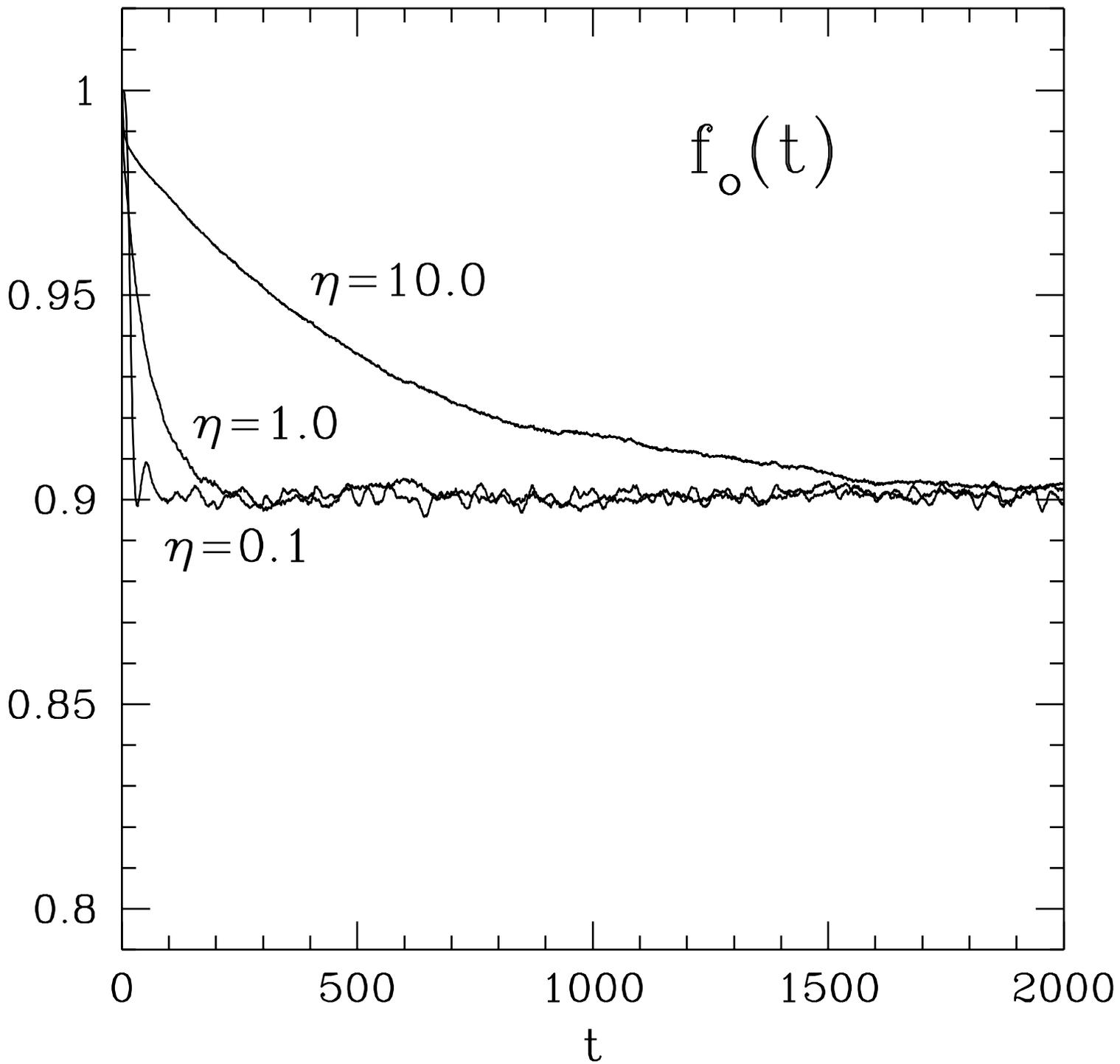
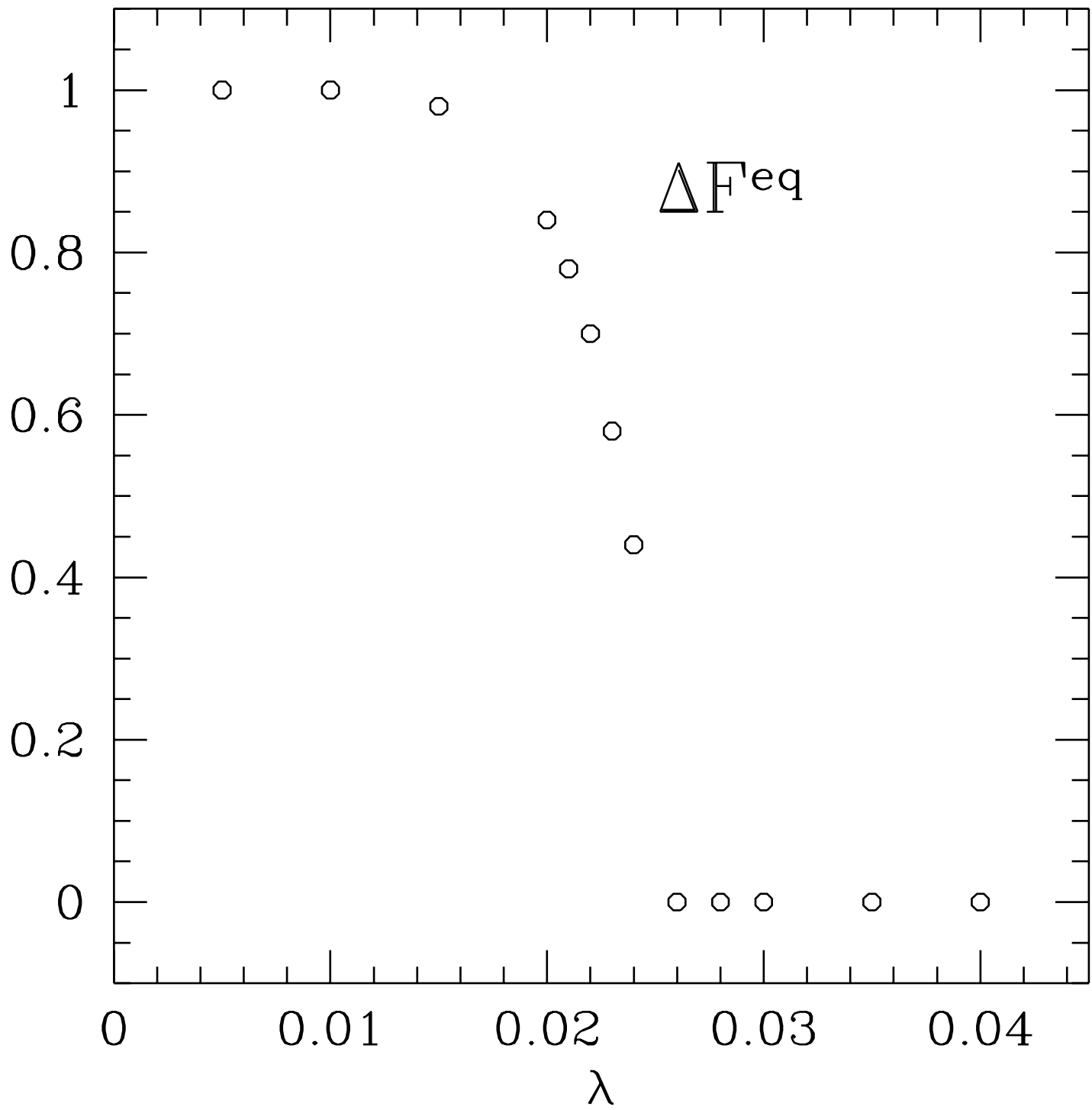
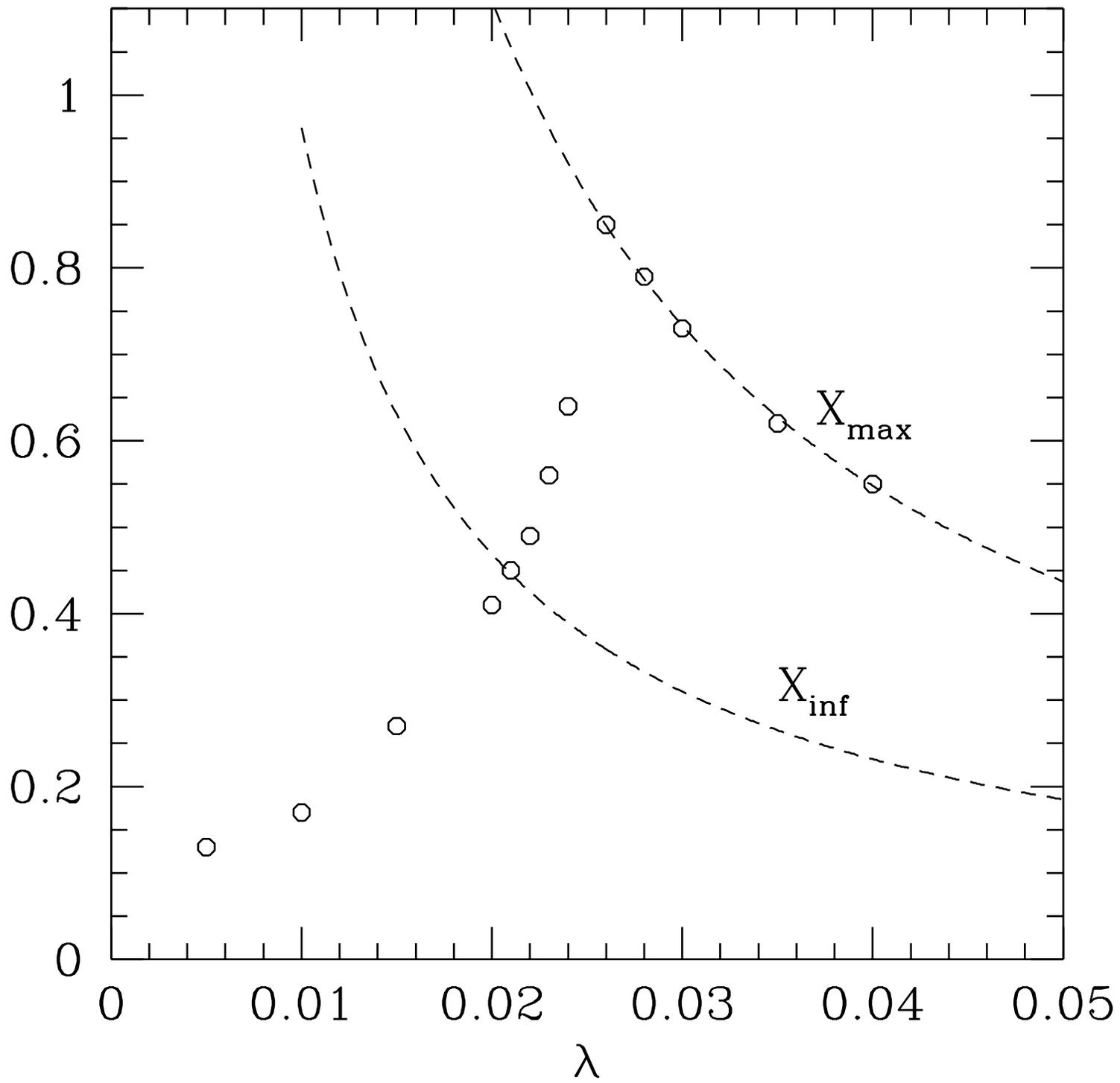
\end{document}